\documentclass[twocolumn,epjc3]{svjour3}  

\RequirePackage[T1]{fontenc}

\smartqed  

\RequirePackage{graphicx}
\RequirePackage{mathptmx} 
\RequirePackage{flushend}
\RequirePackage[numbers,sort&compress]{natbib}
\RequirePackage[colorlinks,citecolor=blue,urlcolor=blue,linkcolor=blue]{hyperref}
\journalname{Eur. Phys. J. C}
\begin{document}
\title{Direct determination of neutron lifetime in
$\beta$-decay}
\author{V.V. Vasiliev 
\\e-mail {
basil\_v@itep.ru}}
\institute{NRC "Kurchatov Institute"-ITEP \\25, Bol. Cheremushkinskaya, Moscow, Russia}
\date{Received: date / Accepted: date}
\maketitle
\begin{abstract}
The neutron lifetime is determined for now most accurately by two methods --the so-called "beam method" and the method of neutron storage in a trap. The goal of this research is to obtain the neutron lifetime with a higher precision by eliminating the traditional systematic errors. 
A new feature is extracting the neutron lifetime only from the set of electron counting rates with their errors. The newly received neutron lifetime value is
$$
\tau_{NT}=883.31\pm0.02(stat.)\pm0.02(sys.) \quad \mbox{s.} 
$$
The obtained neutron lifetime is the weighted average of two lifetimes, defined for different subsets of neutrons providing the well-known effect of the asymmetry in the neutron $\beta$-decay. Simple calculations predict the values of these lifetimes and show their correspondence to the known asymmetry parameters of neutron decay. The arithmetic mean (central) value for the two newly introduced neutron lifetimes was determined. The resulting weighted value $\tau_{NT}$ is in a good agreement with the lifetimes in the first and second methods in limits of their double errors, but exceeds them significantly in precision. In addition, estimates of the new lifetimes, which are called here as $L$-neutron lifetime and $R$-neutron lifetime, are fulfilled. The estimation results correspond to the known parameters of the neutron decay asymmetry.


\end{abstract}

\section*{Introduction}
\label{intro}
Measuring the neutron lifetime is a whole epoch in the history of nuclear physics. 
This era began in the late 1940s and consists of two periods. 
In the first period, the results for the neutron lifetime exceeded 900 seconds. 
The so-called beam method measured a counting rate of protons or
electrons from the decay of neutrons $n \to p+e^-+\tilde \nu $
in a neutron beam of a nuclear reactor ~\cite{SS59}, ~\cite{Chr72}, ~\cite{Brn80}. 
For the second period, typical neutron lifetimes 
were less than 900 seconds. During this period, another method 
of measuring the neutron lifetime becomes the leading one, namely 
the method of storing ultracold neutrons in a trap
until beta-decay. In 2018, this method yielded
the neutron lifetime equal to $881.5 \pm 0.9$ s \cite{Srb2018}. 
The most accurate result of the beam method is 
$\tau_n=887.7 \pm2.2$ s \cite{Yue2013}. 
There were cases of recalculating the results of the
experiments and shifting it from the upper range of values to
the lower one. Herewith, the shift in the estimates of the
lifetime significantly exceeded the indicated experimental
errors. In the case of \cite{SS59} in 1959, the result was 
$1013 \pm26$ seconds and was reduced in 1978 to the 
value of $877 \pm 8$ seconds \cite{BS78} while repeating 
the basic scheme of the experiment.
In the experiment \cite{Brn80} the result of 1980 was $937 \pm 18$
seconds, but 16 years later, the authors published their result
\cite{Brn96} as equal to $889.2 \pm 4.8$ s. The averaging out all the
results for the whole mentioned measurement period without any
restriction leads to an average neutron lifetime of about 900 s. 

The beam experiments use the differential decay equation:
$$
\frac{dN_d}{dt}=\frac{1}{\tau_n} \cdot \varepsilon \times N, \eqno	(1)
$$
where $\frac{dN_d}{dt}$  is the counting rate of the decay
electron (or proton) detector, $\tau_n$  is the neutron
lifetime, $N$  is the number of neutrons at time $t$ in some
region of the neutron beam, $\varepsilon$  is the total
efficiency. The total efficiency includes the efficiency of
neutron detection by a neutron detector, the efficiency of
collecting electrons (or protons) from the decay region to
the electron (proton) detector, and the efficiency of their registration
by the detector. The most difficult problem is the
exact experimental determination of the number $N_d=(\varepsilon \times N)$ 
on the right side of equation (1), i.e. the number of neutrons 
whose decay is in the field of view for the electron (or proton) detector. 
The value $(\varepsilon \times N)$ includes all sources of
systematic errors of the beam method.  

A new method for the beam experiments  has been developed in some years. 
This method excludes necessities to  measure precisely  the number of neutrons 
in the beam, and eliminates the need to determine the absolute values 
of the efficiency components.
\section{ Variation method of neutron decay scale tuning}
\label{sect1:VARM}
The method proposed by the author \cite{VV1} uses a step-wise variation
of the initial number of neutrons passing through a region
controlled by a detector of electrons. The method is based on a system 
of differential equations of type (1) for $k$ steps of 
neutron numbers:
$$
R_d(i)=\frac{1}{\tau_n} \times N_d(i), \eqno	(2)
$$
where $R_d(i)$ is the electron count rate from the detector,
$N_d(i)$ is a number of neutrons seen by the electron detector at
the $i$-th variation step, $i=1, 2, 3,\cdots k$. The problem of measuring 
the number of neutrons is not posed here at all. At each
$i$-th stage of the neutron number variation the count rate
$R_d(i)$ of the electron detector is measured and, after
multiple iterations, the count rate error $\sigma_d(i)$ at each
stage is determined. As a result, an array $(R_d(i),\sigma_d(i))$ 
with $k$ lines is formed. To simplify the notation, it is worth 
eliminating the indices $d$ and $n$ in (2). 
The next step is to represent an unknown set of neutron 
numbers $N_i$ by members of an arithmetic progression 
$N_i\approx \frac{1}{\mu}\times m_i$ with $\frac{1}{\mu}$ 
as a decimal common difference. The common
difference of the required arithmetic progression is the step of
the neutron number scale that describes the distribution of
counting rates with their errors. The integer $m_i$ is the
number of the scale division corresponding to the neutron
number $N_i$. The parameter $\mu$, the inverse of the scale
step, is called a scale factor or $\mu$-factor. 
Then the system of differential equations of decay has the following form:
$$
\tau \times (R_i \pm \sigma_i)\approx \frac{1}{\mu} \times m_i, \eqno	(3)
$$
where $i=1, 2, 3,\cdots k$, $k$ is the full number of
variation steps.

The goal is set as follows. It is necessary to choose an
optimal scale step to describe in the best way the measured data
array of count rates by a certain sample of members of the
obtained arithmetic progression of neutron numbers. 
The $\frac{1}{\mu}$ scale step uniquely identifies the set 
of integers $m_i$ - the scale division numbers corresponding
 to the array of pairs $(R_i,\sigma_i)$ for a given value of $\tau$--trial 
 lifetime. 
The estimate $\aleph_i$ of neutron number $N_i$ is
$$
\aleph_i =
round \left[\frac{round \left[\mu \cdot \tau \cdot R_i,0 \right]}{\mu}, p \right] \mbox{.} \eqno (4)
$$
The operator $round[C, p]$  rounds to the nearest number 
 with $p$ significant digits. The operator $round[C, 0]$ means 
rounding $C$ to the nearest integer. 
The following error functional is constructed for the
required range of $\tau$ for different $p$:
$$
F_{\mu,p}(\tau)=\sum_{i=1}^k \frac{(R_i-\frac{1}{\tau} \cdot
\aleph_i(p,\mu,\tau))^2}{\sigma^2_i} \mbox{.} \eqno (5)
$$
The scale factor $\mu_0$ is to be selected for the best
approximation by the estimate
$\aleph_i(p,\mu_0,\tau)$  for any $p$.
The neutron lifetime $\tau_0$ is determined from the equation
$$
\frac{dF_{\mu_0,p}(\tau)}{d\tau}=0 \mbox{.} \eqno	(6)
$$
\begin{figure}[h!]
\includegraphics{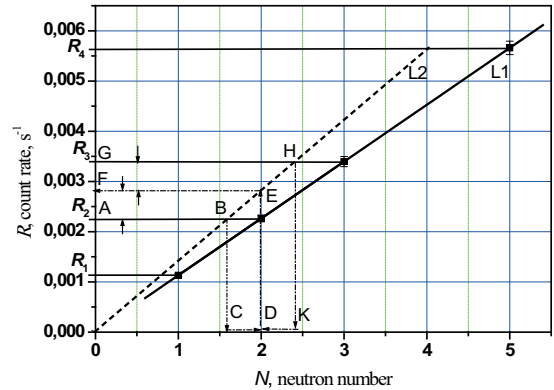}
\caption{Description of count rates with integer neutron numbers.}
\label{fig1}
\end{figure}
Fig. ~\ref{fig1} illustrates the method in the case of integer neutron
numbers. The figure shows the best correspondence of count
rates to the set of neutron numbers 1, 2, 3, 5 by the straight
line L1. The path AB-BC-CD-DE-EF describes the action of the
operator (4) in the case of the count rate $R2$ for the line L2.
At the counting rate $R3$ the operator implements the
trajectory GH-HK-KD-DE-EF. The description of the line L2 in
integers leads to an increase in the error functional by the deviations
FA and GF. Hence a deviation from the optimal line (from L1 to
L2) leads to an increase in the approximation error while
processing in the same neutron number scale. The main
requirement for obtaining an accurate result is a high accuracy
of counting rate measurements.
\section{Experimental data}
\label{sect2:ExpData}
The experimental data of background measurement in the last  ITEP experiment 
on the magnetic storage of ultra-cold neutrons (UCN) was used. 
The background consisted of electrons generated in the vacuum 
chamber of the magnetic trap by the decay of neutrons. 
Electrons  were transported from the trap to the UCN detector \cite{VV2}. 
The proportional gas chamber of the UCN detector operated as an 
electron detector of high efficiency. 
The vertical and horizontal channels of the reactor with open shutters 
were sources of thermal, intermediate and fast neutrons of the neutron 
background. The set of those neutrons penetrating through 
the walls of the trap was the generator of decay electrons. 

\begin{figure}
\includegraphics{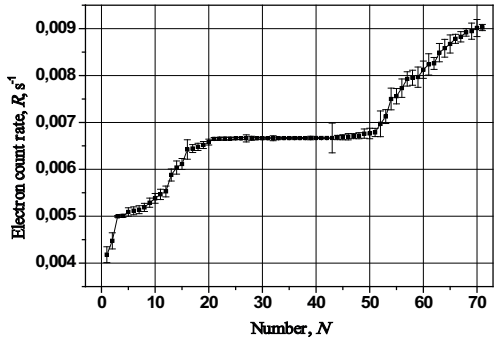}
\caption{Electron count rate vs number. Number 1-71.}
\label{fig2} 
\end{figure}

\begin{figure}
\includegraphics{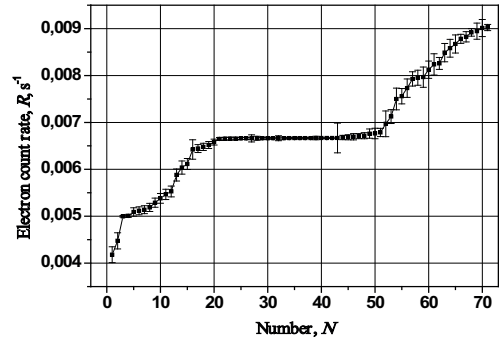}
\caption{Electron count rate vs number.  Number 72-145.}
\label{fig3} 
\end{figure}

\begin{figure}
\includegraphics{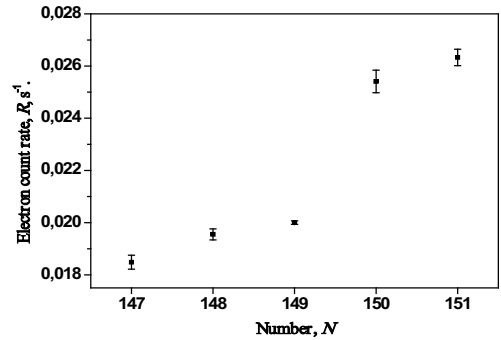}
\caption{Electron count rate vs number.  Number 147-151.}
\label{fig4} 
\end{figure}

To measure the electron background,  a separate long experiment  
was performed.  The low pressure gas  detector was specially 
optimized for counting electrons emanating from the magnetic trap. 
A special absorber was installed into the trap chamber for  
eliminating the ultracold neutrons. A virtually complete storage cycle for electrons 
collected in the trap from background neutrons was carried out. 
The electron counts in the intervals with the magnetic shutter 
on and in the drain intervals with the magnetic shutter off 
were measured. The counting of the electron background 
from the neutron flux through the magnetic trap 
was an analogue of the beam experiment. The count of electrons 
flowing to the  detector from the magnetic trap changed cyclically with 
the changes in the set of simultaneously operating neutron 
channels of the reactor. The data on background measurements in 
the readout intervals of the outgoing electrons were processed 
and shown in the order 
of growth in Fig. ~\ref{fig2}--Fig.~\ref{fig4}.
The full series of 152 count rates is divided into two series.
There are seventy-one values (``series 71'', S-71) in the first
series (Fig. ~\ref{fig2}). The most accurate among these values is
$6.6667\cdot10^{-3} \pm 6\cdot10^{-7}$ $s^{-1}$. In addition,
eighty-one values formed the second series (``series 81'', S-81)
Fig. \ref{fig3} --Fig. \ref{fig4}. Out of 81 measurements,
 for reasons of scale, two points are not shown: 
$$
\mbox{point number} \quad 146: R = 1.355 \cdot10^{-2} \pm 6 \cdot 10^{-5} \quad s^{-1}
$$ 
and 
$$
\mbox{point number} \quad 152: R = 3.738 \cdot10^{-2} \pm 4 \cdot10^{-4}\quad  s^{-1} \mbox{.}
$$
There are several steps of the  background electrons in 
Fig.~\ref{fig2}~--Fig.\ref{fig4}. 
Fig. ~\ref{fig4} shows the count of electrons flowing out of the
magnetic trap after opening the magnetic shutter.
All results  for different reading-out intervals were received
over a period of more than 100 days. The paper \cite{VV2} 
described the scheme and description of the
experimental set-up in more details.

\section{Decay asymmetry and neutron lifetime}
\label{sect3:DAs}
The well-known phenomenon of electron-spin asymmetry of the neutron 
beta decay $n \to p+e^{-}+\tilde \nu$ reveals itself in the fact that neutron 
decays with an electron emitting in the direction of the neutron spin occur 
less frequently than neutron decays with an electron emitting against 
the neutron spin direction. The neutron decay probability 
\cite{Jacks} modified for decay electrons emitting is
$$
dW(E_e,\Theta_e)=W_0dE_ed\Theta_e\left(1+A\cdot\frac{v_e}{c}
\cdot\cos\theta_e\right) \mbox{,} \eqno	(7)
$$
where $A$ is the correlation coefficient of the electron emission with the direction 
of the neutron spin, $\theta_e$ is an electron emission angle 
relative to the direction of the neutron spin, 
$\Theta_e$ is a solid angle of electron emission, $\frac{v_e}{c}$  is an electron helicity, 
and $W_0$ is a constant. From numerous experiments  \cite{PDG} it is known 
that the coefficient $A = - 0.1173 \pm 0.0013$. Thus, neutron decays with electron 
emission in the direction of the neutron spin and against the spin differ qualitatively and 
quantitatively. The decay asymmetry in case of a transversely polarized neutron beam 
is the relative difference of the electrons emitted in the direction of the neutron spin 
and against the direction of the neutron spin. However, the fact that even with 
a completely depolarized ensemble of neutrons, the asymmetry of decay leads 
to two different frequencies of electron generation in the decay region remains 
unnoticed. Nevertheless, the asymmetry of neutron decay is a phenomenon of 
existence of two $\beta$-decay constants, i.e. two reduced decay frequencies 
defined at different subsets of neutrons. The total set $T$ of neutrons is a sum 
of two subsets. Those are the subset of $L$-neutrons with decays by an electron 
ejection against the direction of the neutron spin ($L$-channel) and the subset 
of $R$-neutrons, decaying with the emission of an electron in the direction of 
the neutron spin ($R$-channel). Hence $T=L\cup R$ and the number of neutrons 
$N_T$ in the total set  $T$ is the sum of $L$-neutrons ($L$-subset) and $R$-neutrons 
($R$-subset): $N_T=N_L+N_R$. Without loss of generality the decay constants 
for these neutron sets are equal to
$$
\lambda_S=\frac{1}{N_{S}}\cdot\frac{dN_{S}}{dt}\mbox{,}  \eqno (8)
$$
where $S=T, L, R$.
Differentiation of the sum $N_T$ in the expression (8) for $S=T$ with the 
subsequent insularity of the partial constants for $S=L$ and $S=R$ 
leads to the expression: 
$$
\lambda_T=\lambda_L \cdot W_L+ \lambda_R \cdot W_R \mbox{.} \eqno (9)
$$
Here the total decay constant $\lambda_T$ has the form of a weighted average 
of the partial constants $\lambda_L$ and $\lambda_R$. 
The weights $W_L$ and $W_R$, when use the parallel decay rule \cite[ p. 344]{Blatt} as  
$\frac{N_R}{N_L}=\frac{\lambda_R}{\lambda_L}$  
are the following relations
$$
W_L=\frac{N_L}{N_L + N_R}=\frac{\lambda_L}{\lambda_L  + \lambda_R} \mbox{,} \eqno (10.1)
$$
$$
W_R=\frac{N_R}{N_L + N_R}=\frac{\lambda_R}{\lambda_L + \lambda_R} \mbox{.} \eqno (10.2)
$$
The lifetime of neutrons in every $S$-set is 
$\tau_{NS} = \frac {1} {\lambda_S}$
and the total lifetime on the full set $T$ is equal to 
$\tau_{NT} = \frac {1} {\lambda_T}$. 
Therefore, a record for the total lifetime follows from (9) in the form of 
a weighted average value, namely:
$$
\tau_{NT}=\tau_{NL} \cdot W_{\tau L}+\tau_{NR} \cdot W_{\tau R}\mbox{,} \eqno (11)
$$
where the weights for the partial lifetimes receive the following forms:
$$
W_{\tau L} = \frac{\tau^2_{NR}}{\tau^2_{NL}+\tau^2_{NR}}\mbox{,}	\eqno (12.1)
$$
$$
W_{\tau R} = \frac{\tau^2_{NL}}{\tau^2_{NL}+\tau^2_{NR}}\mbox{.}	\eqno (12.2)
$$
The introduced partial decay constants receive the following dependence on 
the asymmetry parameter $\Delta$
$$
\lambda_L=\lambda_0\cdot (1+\Delta) \mbox{,} \eqno (13.1)
$$
$$
\lambda_R=\lambda_0\cdot (1-\Delta) \mbox{,}	\eqno (13.2)
$$
where $\lambda_0$  is the arithmetic mean of the introduced constants, 
$\Delta=A\cdot\frac{\bar v_e}{c}$ , where $\frac{\bar v_e}{c}$ is the 
average helicity of electrons emitted during neutron decay. 
The average helicity of electrons can be calculated from the electronic 
neutron decay spectrum, for example, from the results of \cite{Robsn}. 
The partial lifetimes are 
$$
\tau_{NL}=\tau_{NCenter} \cdot (1-\Delta) \mbox{,} \eqno (14.1)
$$
$$
\tau_{NR}= \tau_{NCenter} \cdot (1+\Delta) \mbox{,} \eqno (14.2)
$$
where $\tau_{NCenter}$ is a central neutron life time , i.e. the central 
point between values of $\tau_{NL}$ and $\tau_{NR}$ .
The weighted average decay constant $\lambda_T$  is related to the average 
decay constant $\lambda_0$  as follows:
$$
\lambda_T=\lambda_0 \cdot (1+\Delta^2) \mbox{.} \eqno (15)
$$
The total neutron lifetime $\tau_{NT}$ vs $\Delta$ is
$$
\tau_{NT}=
\tau_{NCenter} \cdot \left(1-\frac{2 \cdot \Delta^2}{1+\Delta^2}\right) \mbox{.} \eqno (16)
$$
Therefore, the dependence of the error functional (5) on the trial lifetime is symmetrical with respect to the point $\tau_{NCenter}$. Then the weighted average lifetime according to (16) is to the left of the center of symmetry. 
The goal of this study, therefore, turned out to be two-fold. 
Firstly, it is a direct determination  of an experimental value for the observed neutron lifetime, i.e. the weighted average neutron lifetime.  
On the other hand, it became necessary to determine the value of the ``central'' neutron lifetime. In the case of a convincing correspondence between a determined displacement value and a calculated value (16), it is advisable to introduce new physical quantities into the physical dictionary -- the lifetimes of $L$-neutrons and $R$-neutrons and to estimate their numerical values using formulas (14.1) and (14.2). 

\section{Direct determination of the neutron lifetime from the experimental data}
\label{sect4:DDED}
\begin{figure}
\includegraphics{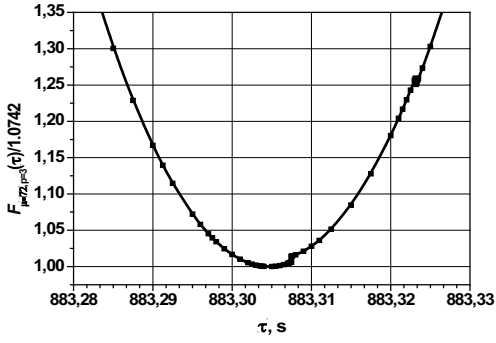}
\caption{ Error functional dependence on trial lifetime for S-152.}
\label{fig5} 
\end{figure}

\begin{figure}
\includegraphics{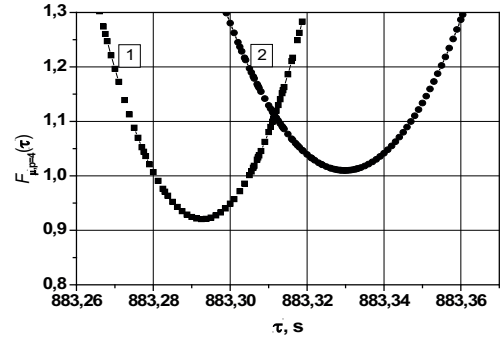}
\caption{ Error functional dependence on trial lifetime  for S-71 and S-81.}
\label{fig6} 
\end{figure}

\begin{figure}
\includegraphics{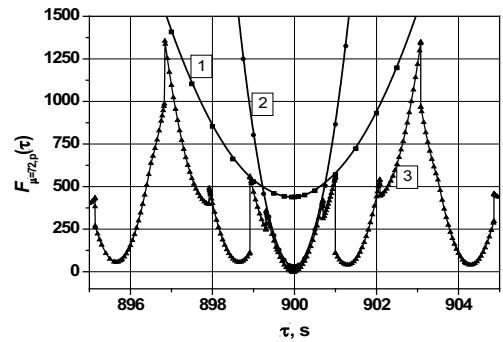}
\caption{Center of symmetry of the error functional for S-152 at $\mu=72$}
\label{fig7} 
\end{figure}

The interval of the trial lifetime from 860 seconds to 940 seconds is quite
 informative for solving the given problem. This interval includes 
the results of measurements of the lifetime of the last three decades. 
To determine the neutron lifetime, it is necessary to find the minimum value 
of the scale-factor $\mu$, providing the condition of ``reduction-to-one''. 
This condition means that  the minimum point of the error 
functional is the closest to the unity for any data series in the considered
range of the lifetime. 
The ``reduction-to-one'' is applied to the data shown in Fig. ~\ref{fig2}--Fig.~\ref{fig4} 
for various values of the scale step in the indicated range of the trial 
neutron lifetime $\tau$. The results of the data processing  
are illustrated by Fig. ~\ref{fig5}--Fig.~\ref{fig7}. 

Fig. ~\ref{fig5} shows the result of the reduction-to-one 
for the S-152 series with  $p=3$. The closest to the unity is 
the value of the functional minimum equal to 1.074, corresponding to
 the lifetime of 883.305 s for the scale-factor $\mu =72$. At the same 
scale-factor $\mu$, minimums of the functional of the second, third and 
fourth orders of accuracy reached a good agreement. The value for 
the neutron lifetime is 
$\tau_{NT}=(529983\pm10)\cdot\frac{1}{6}\cdot10^{-2}$ s  
after using the half-width of the parabola of the third order functional 
at height $\chi^2=1.2$. 

Fig. ~\ref{fig6} shows the results of the reduction-to-one for 
the S-71 (1) and S-81 (2) series separately. The dependence of the average 
weighted functional over two series on the lifetime gives the result for
the neutron lifetime $\tau_{NT}=(264992\pm7)\cdot\frac{1}{3}\cdot10^{-2}$ s.
Thus, the complete result for the neutron lifetime is 
$\tau_{NT}=883.31\pm0.02$ s, 95\%, $CL$.
 
All these facts prove the following:

a) both independent data series S-71 and S-81 are qualitatively homogeneous
 and describe electrons from the decay of neutrons without admixture 
of an additional background;

b) the results of independent data series S-71 and S-81 are compatible 
within statistical errors, reasonably processed together and do not require 
an additional introduction of a systematic error to describe the consolidated result.

Therefore, the applied method has reached its goal of eliminating known 
systematic errors for higher accuracy in the experimental determination of 
the neutron lifetime.
  
Additional studies were made to prove the existence of a common center of 
symmetry of the error functional on the trial lifetime interval 
from 895 to 905 seconds for all orders of accuracy $p = 0, 1, 2, \cdots $.

Fig.~\ref{fig7} shows results for the scale-factor $\mu=72$  by dependencies
 (1)-(3) of error functional $F_{\mu,p}(\tau)$:  (1) integers ($p = 0$); (2) $p=1$; (3) $p=2$. 
The coincidence of the minimums of all three orders of the functional on the 
$\tau$-scale at the point of 899.975 s, with symmetry of the left and right wings 
of the functional shows that this point is the center of symmetry and confirms 
the result of \cite{VV2}. This value can serve as an estimate of 
the so-called ``central'' neutron lifetime. 
The symmetry of the coordinates of the jumps of the functional (3) with respect 
to this $\tau_{NCenter}=899.975$  s point also confirms its central position. 
The dependence of the functional on the trial lifetime allows determining 
the value of the central neutron lifetime with a completely moderate error 
$\tau_{NCenter}=(179995\pm 4)\cdot 5 \cdot 10^{-3}$ s. It is important 
to emphasize that the error functional in the range of the trial lifetime of 
750-1050 s has only one specified center of symmetry for all accuracy level $p$.

\section{Upper limit of a systematic error}
\label{sect5:SYS}
As a source of systematic error, a hypothetical difference between the electrons 
flowing through the closed magnetic shutter and the electrons flowing onto 
the detector through the open magnetic shutter may appear. If only the electrons 
of the first type are presented in the first S-71 series, then in the second S-81 
series there is a fraction of electrons of the second type providing 
the maximum values of the counting rate. 
The result indicated in Fig.~\ref{fig6} proves that the independently 
obtained lifetime values in these series are mutually compatible within the 
statistical error. In the case of functional (1), the result is $883.29 \pm 0.03$ s, 
and in the case of functional (2), the result is $883.33 \pm 0.03$ s. 
The errors indicated here are equal to the half-width of the parabolas at the level $\chi^2=1.3$. 
It means that the displacement is within the limits of the statistical errors 
at 95$\% CL$. 
Calculation of the half-sum of these lifetimes and its error gives 
$\tau_{NT}=883.31\pm 0.02$ s, and comparison  with the result 
shown in Fig. ~\ref{fig5} confirms complete coincidence. Moreover, 
the minimums of the functional for S-152, $\mu=72$ at $p=2$, 
$p=3$ and $p=4$ coincide in those limits altogether. 

Nevertheless, the fact that the minimums at the accuracy $p=4$
shift from 883.29 s for S-71 to 883.33 s for S-81 gives 
a reason to interpret it due to some unaccounted factors and to introduce 
on this basis a systematic error equal to the half of the bias value. 
Thus, the estimate of the systematic error is 0.02 s, and the result for 
the neutron lifetime as a weighted average is 
$\tau_{NT}=883.31 \pm 0.02(stat.)\pm 0.02(sys.)$ s, $95\% CL$. 
In the same format, the central neutron lifetime has got the following 
value $\tau_{NCenter}=899.98 \pm 0.02(stat.) \pm 0.02(sys.)$ s. 

\section{Discussion of the result} 
\label{sect6:DISR}
In details the previous two results in the introduction are 
$$
887.7 \pm 1.2 (stat.) \pm 1.9 (syst.) \quad \mbox{s}
$$  for the beam method, 
and 
$$
881.5 \pm 0.7 (stat.) \pm 0.6 (syst.) \quad \mbox{s}
$$  
for the storage method. 
Taking into account limits within double total errors for these two results, 
it is easy to note the coincidence of the result of this work with the lower limit 
for the first result (883.3 s) and with the upper limit for the second result (883.3 s) . 
Therefore, a good agreement between these three results, including the presented 
one, is evident. 

It also means that there are no grounds for any conclusions about the 
so-called "neutron anomaly" as a possible interpretation of the difference 
between the two mentioned results of measuring the neutron lifetime.

A side and unexpected result of the present research is the above evidence of 
the difference between the two obtained  neutron lifetimes, the first of them 
is the weighted average value and the second is the "central" lifetime of neutron. 
Within the error limits (0.02 s), the obtained value 
of the weighted average neutron lifetime in a more convenient form is 
$$
\tau_{NT}=900 \cdot \frac{53}{54} \quad \mbox{s,}
$$ 
while the central lifetime is $\tau_{NCenter}\approx 900$ s within the same error. 
The value of the relative shift 
$$
\delta=\frac{\tau_{NCenter} - \tau_{NT}}{\tau_{NCenter}+\tau_{NT}}
$$
is a characteristic of the displacement of the weighted average lifetime relative to 
the central lifetime. Then the estimate of the relative shift is 
$$
\delta=\frac{1}{107} \mbox{.}
$$ 
The weighted average lifetime is connected with the central lifetime as 
$$
\tau_{NT}=\tau_{NCenter} \cdot \left(1-\frac{2 \cdot \delta}{1+\delta}\right) \mbox{.}
\eqno		(17)
$$
From (16) $\delta$ is interpreted as $\delta=\Delta^2$. 
The obtained value for $\Delta=\frac{1}{\sqrt{107}}$ is in a good agreement with $A = - 0.1173 \pm 0.001$ 
and $\frac{\bar v_e}{c}=0.824$. This value for the average helicity is confirmed 
by the spectrum of electrons from the neutron decay \cite{Robsn}. 

Thus, the displacement of the weighted average neutron lifetime relative to the 
so-called central lifetime is a consequence of the electron-spin asymmetry of neutron decay. 
The main purpose of this research is the direct determination of the observed neutron lifetime. 
Nevertheless, it is easy to predict from the displacement the numerical values for 
the lifetime of $L$-neutrons $\tau_{NL}$  (14.1) and
the lifetime of $R$-neutrons $\tau_{NR}$  (14.2). They are the following:
$$
\tau_{NL}=(2 \cdot 3 \cdot 5)^2 \cdot \left (1-\frac{1}{\sqrt{107}} \right)  \quad \mbox{s,}
$$ 
$$
\tau_{NR}=(2 \cdot 3 \cdot 5)^2 \cdot \left (1+\frac{1}{\sqrt{107}} \right)  \quad \mbox{s.}
$$ 
The errors of the indicated values are estimated not exceeding 0.08 s. 
A direct determination of these parameters of the neutron ~$\beta$-decay will make 
it possible in the future to indicate their values more accurately. 

Now it is worth noticing that the weighted average of the mentioned above 
values of $\tau_{NL}$ and $\tau_{NR}$ is in a good agreement with 
the neutron lifetime obtained in this research from the experimental data. 
This can be easily verified using numerical expressions for the lifetime weights of 
(12.1) and (12.2), namely:
$$
W_{\tau L}=\frac{1}{2}+\frac{\sqrt{107}}{108} \mbox{,}
$$
$$
W_{\tau R}=\frac{1}{2}-\frac{\sqrt{107}}{108} \mbox{.}
$$
The calculation result by formula (11) gives 883.33 s 
that coincides with the experimental result of this work within $\pm 0.01$ s.

\section{Conclusion} 
\label{sect7:CONCL}
Using the proposed method, the neutron lifetime is  determined with 
the accuracy of 0.02 s. The system of differential equations of decay 
with a step-wise variation of the number of neutrons turned out 
to be a sufficient tool for determining the neutron lifetime  by 
the modified least-squares method. 
Traditional sources of systematic errors inherent in the beam method of measuring 
the neutron lifetime are successfully excluded.  As a result, the accuracy of the neutron lifetime  is significantly improved. The observed (weighted average) neutron lifetime obtained in this work is expressed in primes as 
$$
\tau_{NT}=2 \cdot 5^2 \cdot \frac{53}{3} \quad \mbox{s}
$$  
within no more than 0.03 s (taking into account the introduced systematic error).

In addition to the main result, estimates of the effect of splitting the neutron lifetime 
are obtained.  It is indicated that neutron decay can be described by two lifetimes: 
$L$-neutron lifetime and $R$-neutron lifetime, differing in the sign of the scalar 
product of the electron momentum and the neutron spin. The estimation results 
correspond to the known parameters of the electron-spin asymmetry of the neutron decay. 
This fact provides the grounds for introducing new quantities -  $L$-neutron lifetime 
and the $R$-neutron lifetime into the neutron ~$\beta$ -decay physics. 
The obtained interval of neutron lifetimes from $\tau_{NL} = 813$ s to 
$\tau_{NR} = 987$ s gives an explanation of the range of experimental values 
obtained over the 70-year history of of the neutron lifetime measurements. 
This method opens up the prospect of reducing the neutron lifetime errors 
to thousandths of a second.

\begin{acknowledgements}

The author is deeply grateful to V.V. Vladimirsky  for his personal support. 
The author is grateful to the staff of the UCN-ITEP group  - V. F. Belkin, A. A. Belonozhenko, N. I. Kozlov, E. N. Mospan, A. Yu. Karpov, I. B. Rozhnin, A.M. Salomatin , V.G. Frankovskaya, and also and especially  G.M. Kukavadze and S.P. Boroblev for the creation and operation of the KENTAVR set-up at the ITEP HWR reactor.
\end{acknowledgements}

\end{document}